\documentclass[twocolumn,aps,pra,preprintnumbers,amsmath,amssymb,superscriptaddress,floatfix,longbibliography]{revtex4-2}

\usepackage{latexsym}
\usepackage{layouts}
\usepackage{graphicx}
\usepackage{textcomp}
\usepackage{color}
\usepackage{mathtools}
\usepackage{lipsum}

\usepackage{float}
\makeatletter
\let\newfloat\newfloat@ltx
\makeatother

\usepackage{algpseudocode,algorithm,algorithmicx}
\def\ALG@special@indent{%
    \ifdim\ALG@thistlm=0pt\relax
        \hskip-\leftmargin
    \else
        \hskip\ALG@thistlm
    \fi
}

\begin{document}
\title{Channel-noise tracking for sub-shot-noise-limited receivers with neural networks}
		
\author{M. T. DiMario}
\affiliation{Center for Quantum Information and Control, Department of Physics and Astronomy, University of New Mexico, Albuquerque, New Mexico 87131, USA}

\author{F. E. Becerra}
\affiliation{Center for Quantum Information and Control, Department of Physics and Astronomy, University of New Mexico, Albuquerque, New Mexico 87131, USA}
\email{fbecerra@unm.edu}

\begin{abstract}
Non-Gaussian receivers for optical communication with coherent states can achieve measurement sensitivities beyond the limits of conventional detection, given by the quantum-noise limit (QNL). However, the amount of information that can be reliably transmitted substantially degrades if there is noise in the communication channel, unless the receiver is able to efficiently compensate for such noise. Here, we investigate the use of a deep neural network as a computationally efficient estimator of phase and amplitude channel noise to enable a reliable method for noise tracking for non-Gaussian receivers. The neural network uses the data collected by the non-Gaussian receiver to estimate and correct for dynamic channel noise in real-time. Using numerical simulations, we find that this noise tracking method allows the non-Gaussian receiver to maintain its benefit over the QNL across a broad range of strengths and bandwidths of phase and intensity noise. The noise tracking method based on neural networks can further include other types of noise to ensure sub-QNL performance in channels with many sources of noise.
\end{abstract}

\maketitle

\begin{figure*}[t]
	\includegraphics[width = 0.95\textwidth]{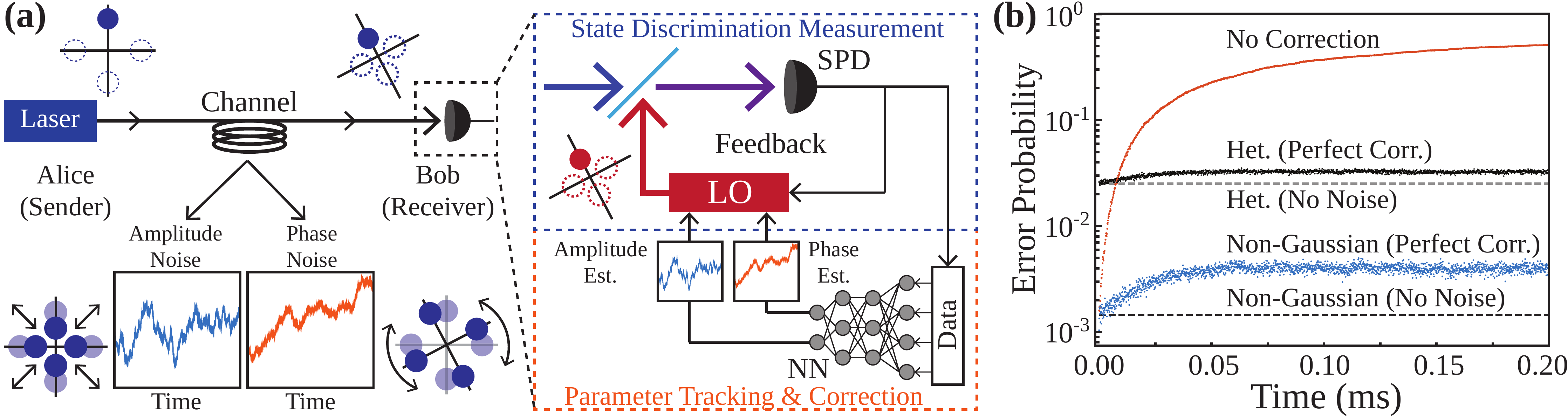}
	\caption{\textbf{Channel noise tracking method.} (a) Schematic of two parties, Alice and Bob, communicating over a noisy channel with Bob at the receiver performing phase and amplitude noise tracking. The receiver uses an adaptive non-Gaussian measurement strategy to realize state discrimination below the QNL using interference between the input state and a local oscillator (LO) followed by photon counting with a single photon detector (SPD). A neural network (NN) takes the data collected from the non-Gaussian receiver and outputs estimates for the current phase offset and input intensity. The estimates are fed-forward to the LO to match the input intensity and counteract phase noise in real-time. (b) Effect of channel noise with (blue) and without (orange) perfect parameter tracking (see main text) for the non-Gaussian receiver. Black points show a perfectly-corrected heterodyne receiver, and black and gray dashed lines show the error for a non-Gaussian and a heterodyne receiver in the absence of noise, respectively.}
	\label{strategy}
\end{figure*}

\section{Introduction}
The intrinsic properties of coherent states can enable efficient and practical classical \cite{giovannetti04,li09,ip08,kikuchi08} and quantum \cite{arrazola14,ghorai19,pirandola17,gisin02,gisin07} communications. When utilizing the phase of coherent states combined with their intensity to encode and transmit information, higher rates of information transfer may be achieved compared to communication schemes using intensity-only encodings \cite{kikuchi08,kikuchi16}. However, channel noise can severely limit the advantage of communications with coherent encodings. In conventional coherent communications, the optical receiver performs a heterodyne measurement with  shot-noise-limited sensitivity, corresponding to the quantum-noise limit (QNL). This measurement allows for the use of post processing methods of the collected data to estimate channel noise and correct the data to recover the transmitted information \cite{jouguet03,armada98,marie17,kikuchi16,qi15,soh15,wang19,ip07,lygagnon06,bina16}. While current coherent optical communications rely on these conventional approaches, a heterodyne measurement cannot reach the ultimate limit of sensitivity \cite{helstrom76} and information transfer \cite{giovannetti04,giovannetti14,banaszek20}.

In contrast to conventional strategies, non-Gaussian receivers can surpass the QNL, providing higher measurement sensitivities for decoding information \cite{mueller12,mueller15,becerra13,becerra15,lee16,ferdinand17,izumi12,izumi20b}. However, in the presence of channel noise, the benefit of non-Gaussian receivers over conventional strategies critically depends on the ability to perform efficient channel-noise tracking. Recent work demonstrated an efficient method for phase tracking for non-Gaussian receivers \cite{dimario20}. This phase tracking method estimates and corrects for the phase noise in real time, which is required by the strategies used in non-Gaussian receivers, as opposed to post processing of the collected data with heterodyne detection. This method enabled sub-QNL sensitivity in the presence of phase noise, which is particularly damaging for coherent encodings \cite{bina17,dimario20}.

In more realistic situations there may be multiple sources of noise present in the communication channel, such as thermal noise \cite{habif19,yuan20}, phase diffusion \cite{dimario19,genoni11,genoni12}, phase noise, and amplitude noise. In such situations the non-Gaussian receiver must perform efficient high-dimensional parameter estimation and tracking in order to maintain the expected sub-QNL performance. However, current methods for single-parameter noise tracking cannot be efficiently scaled to higher dimensions for tracking and correction of multiple sources of noise. Thus, enabling noise tracking for non-Gaussian receivers in channels with complex and dynamic noise requires novel and efficient methods for multi-parameter estimation that scale favorably to higher dimensions. Practical parameter tracking also requires estimation on a time-scale which is very small ($\ll1\%$) compared to the bandwidth of the channel noise. For example, realistic kilohertz scale phase noise \cite{lygagnon06,ip07,kikuchi08} would require estimation on at least megahertz time scales, and a Bayesian estimator may not be compatible with this requirement.

Machine learning has been shown to be a powerful tool for solving many problems in coherent communications where conventional methods may be inefficient or computationally difficult \cite{carleo19,wallnoefer19,khan19,mata18,chen19}. In particular, artificial neural networks \cite{schmidhuber15} have seen broad applications in quantum information \cite{dunjko18, melnikov18, hentschel10, lumino18, fiderer20, cimini19, giordani20, lohani18, steinbrecher19, beer20} and optical communications \cite{thrane17, wu09, zibar16, lohani19, karanov18}, and for channel noise estimation and monitoring \cite{zibar15, khan17, wang17}. While machine learning techniques benefit current communication technologies, their application for parameter tracking for non-Gaussian receivers with sub-shot-noise limited performances have yet to be investigated.

In this work, we numerically investigate a method for multi-parameter channel noise tracking based on a neural network (NN) estimator for a non-Gaussian receiver with sub-QNL sensitivity for state discrimination of quaternary phase-shift-keyed (QPSK) coherent states. We construct a NN as a precise and computationally efficient multi-parameter estimator for tracking the time-varying phase and intensity of the input coherent states, and benchmark its performance against a Bayesian estimator, which is expected to be accurate but is computationally expensive to calculate. We find that, across a broad range of channel noise strengths and input powers, the NN-based method for noise tracking shows similar performance to a Bayesian-based noise tracking approach, and allows the non-Gaussian receiver to maintain sub-QNL sensitivity. This shows that a NN estimator is a viable method for real-time, multi-parameter channel noise tracking in non-Gaussian receivers due to its efficiency and potential scalability to higher dimensions. In Sec. II we describe the non-Gaussian receiver strategy and the NN estimator used for the noise tracking method. In Sec. III we investigate the performance of the channel noise tracking. We discuss the results of the work in Sec. IV.
\section{Receiver and estimation strategy}
We numerically study the use of a NN-based method for noise parameter tracking for non-Gaussian receivers based on adaptive measurements and photon counting. As a proof-of-concept, we investigate a NN-based method for tracking phase and amplitude channel noise, that uses only the data collected during the state discrimination measurement. This NN-based method can be easily extended to perform higher dimensional parameter estimation for tracking additional sources of noise in the channel such as thermal noise \cite{habif19,yuan20} or phase diffusion \cite{dimario19,genoni11,genoni12}. In this section, we describe (A) the measurement strategy of the non-Gaussian receiver for coherent state discrimination; (B) how the data from the measurement is used by the NN; and (C) the NN estimator, which can be used for estimation of channel noise from multiple sources.
\subsection{State Discrimination Measurement}
Figure \ref{strategy}(a) shows a scenario where a receiver attempts to perform coherent state discrimination with an adaptive photon counting measurement with sensitivity below the QNL \cite{becerra13,becerra15}. Dynamic phase and amplitude noise induced by the communication channel degrades the attainable sensitivity of the receiver. Tracking the phase and amplitude noise of the input states induced in the channel using the data collected during the discrimination measurement can in principle allow the receiver to correct its strategy and maintain sub-QNL sensitivity.

Here, we study a method for channel noise tracking for a receiver based on an adaptive non-Gaussian strategy \cite{becerra15} for phase coherent states $|\alpha_{k}\rangle\in\{|\alpha e^{i2\pi k/M} \rangle \}$, where $k=0,1,...,M-1$. For $M=4$, this corresponds to quaternary phase-shift-keyed (QPSK) coherent states. The state discrimination strategy consists of $L$ adaptive measurement steps. Each step performs a hypothesis test of the input state using a local oscillator (LO) to implement a displacement operation $\hat{D}(\beta)$ through interference and single photon counting. In each adaptive step $j=1, 2, ... L$, the receiver attempts to displace the most likely state to the vacuum state by adjusting the LO phase $\mathrm{arg}(\beta) = \theta_{j}\in \{0, \pi/2, \pi, 3\pi/2\}$ with $|\beta|=|\alpha_{k}|$, followed by single photon detection. The detector has a finite photon number resolution (PNR) where up to $m$ photons can be resolved, denoted as PNR($m$), before becoming a threshold detector \cite{becerra15}. At the end of the $L$ adaptive steps, the best guess of the receiver $\theta_{\mathrm{disc}}$ for the true input phase is the state with maximum \textit{a posteriori} probability given the entire detection history. As described in Section II(B) and Section II(C), the photon counting data from the adaptive measurement steps together with $\theta_{\mathrm{disc}}$ allows the receiver to perform phase and amplitude tracking, where estimates of the channel noise are fed-forward to the LO in order to maintain the sub-QNL performance of the receiver.

Figure \ref{strategy}(b) shows an example of the error probability for the adaptive non-Gaussian receiver for QPSK states for an average input mean photon number of $\langle \hat{n} \rangle_{0} = |\alpha|^{2}=5.0$, which is proportional to the intensity, averaged over 5000 noise realizations and obtained through Monte-Carlo simulations. For all Monte-Carlo simulations in this study, we assume ideal detection efficiency, zero detector dark counts, a photon number resolution of PNR(10), and $L$=10 adaptive steps. To represent a realistic experiment, we use an interference visibility of the displacement operation of 99.7$\%$ \cite{becerra15}. The blue (orange) points show the error probability for the non-Gaussian receiver with (without) perfect noise tracking. Perfect tracking refers to a situation where the receiver has complete knowledge of the time-dependent input intensity and phase noise induced by the channel. The black points show the error of an ideal heterodyne measurement, performing at the QNL, with perfect tracking \footnote{We note that in conventional optical communications, the use of amplification prior to detection can be used to reduce the probability of error. In those situations, the use of non-Gaussian receivers can further reduce the error rates far beyond what could be achieved by heterodyne measurements. For example, given an input pulse power corresponding to $\langle \hat{n} \rangle_0 = 5.0$, the error reduction for a non-Gaussian receiver ($P_{\mathrm{E,NG}}$) compared to the heterodyne receiver at the QNL equals a factor of QNL$/P_{\mathrm{E,NG}}=17$. At this initial power, the heterodyne receiver would require ~3 dB of noiseless gain to reach the non-amplified non-Gaussian receiver. On the other hand if the same 3 dB amplifier is used with the non-Gaussian receiver, and compared to the amplified heterodyne receiver, then this ratio grows to QNL$/P_{\mathrm{E,NG}}=210$}. The dashed lines show the expected error in the absence of noise for a heterodyne (gray) and non-Gaussian (black) receiver. The error for the non-Gaussian measurement remaining below the heterodyne limit (QNL) shows that if the receiver can implement accurate parameter tracking, then its benefit over the QNL can be maintained. Furthermore, any tracking method for the non-Gaussian receiver requires correcting for dynamical noise in real-time to ensure sub-QNL performance \cite{dimario20}, in contrast to methods for heterodyne receivers, where estimation and correction can be done in post-processing of the data.

\subsection{Detection Matrix}
The measurement data collected by the non-Gaussian receiver from the discrimination of $N$ input states is used for parameter estimation. For the discrimination of one input state, this data consists of the $L$ photon detections $\{ d_{j} \}_{L}$ and relative phases $\{ \Delta_{j} \}_{L}$ between the LO and input state for each adaptive step $j$. Due to the low error rate achieved by the non-Gaussian measurement, the guess $\theta_{\mathrm{disc}}$ of the phase of the input state corresponds to the true input phase with high probability. Thus, $\theta_{\mathrm{disc}}$ can be used to infer the relative phase $\Delta_{j}$ between the LO and actual input state at every adaptive measurement step $j$ such that $\Delta_{j} = \theta_{j} - \theta_{\mathrm{disc}}$, as in \cite{dimario20}. This state discrimination data $\{ \Delta_{j}, d_{j} \}$ is binned into what we refer to as the detection matrix $\textbf{D}$, which is a $M\times(m+1)$ matrix, where $m$ is the PNR of the receiver. After each measurement, the matrix elements $\mathrm{D}_{l,k}$ are incremented by the total number of times that the number of detected photons in an adaptive step $j$ was $d_{j}=l$ and the relative phase was $\Delta_{j}=2\pi k/M$ for $k \in \{0, 1, ..., M-1\}$. Thus, the rows of the matrix $\textbf{D}$ represent the photon number distributions for different relative phases $k\pi/2$ between the LO $(\theta_{j})$ and final hypothesis $(\theta_{\mathrm{disc}})$ for QPSK states \cite{dimario20}. After completing $N$ experiments, the matrix $\textbf{D}$ contains $N\times L$ pairs $\{d_{j}, \Delta_{j}\}$ and it is used for parameter estimation. Once estimation has been performed, the matrix is reset such that $\mathrm{D}_{l,k}=0$ for all $l$ and $k$.
\begin{figure}[!t]
	\includegraphics[width = 8.5cm]{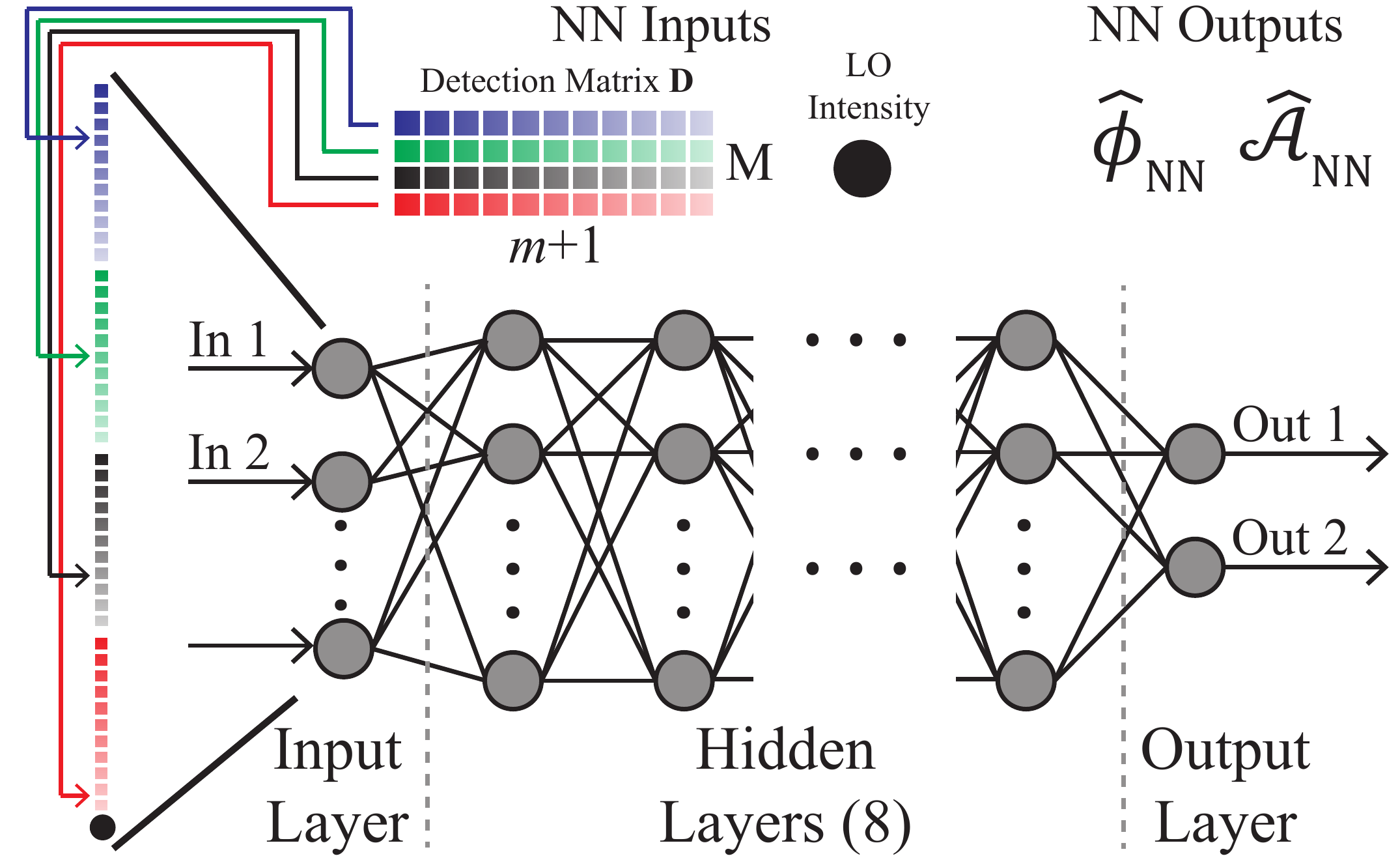}
	\caption{\textbf{Neural network}. The neural network (NN) for noise estimation has 10 layers (8 hidden) with sizes described in Table \ref{nn_param} in Appendix A. The NN inputs are a flattened version of the detection matrix \textbf{D} normalized across each row, and the LO intensity $\mathcal{B}$ for the measurements whose data is contained in the detection matrix. The outputs of the NN are estimates for the phase offset $\hat{\phi}_{NN}$ and input intensity $\hat{\mathcal{A}}_{NN}$.}
	\label{nn_fig}
\end{figure}
In order to extract information from $\textbf{D}$ to correct for channel noise affecting the measurement, the receiver must utilize a particular estimator. A Bayesian estimator, which uses the full likelihood functions, will yield estimates for the channel noise with small uncertainty \cite{lehmann98}. However, this estimator is computationally demanding to calculate. Since the estimation and correction of the channel noise for non-Gaussian receivers must be performed in real-time, a Bayesian method may be incompatible with applications requiring high bandwidth sub-QNL receivers. Therefore, to enable practical implementations of non-Gaussian receivers requires an estimator that is both precise and computationally efficient while being easily scalable to higher dimensions to track multiple sources of channel noise. For example, the simple case of single parameter estimation for phase tracking for non-Gaussian measurements has been experimentally demonstrated \cite{dimario20} using a simple estimator, which is calculated in real-time with minimal computational resources.
\subsection{Neural Network Estimator}
We construct a NN as a multi-parameter estimator which maps the data collected from the state discrimination measurement to estimates for the input intensity and phase offset. We compare the performance of the NN estimator to a Bayesian estimator. The Bayesian-based method for noise tracking serves as a benchmark and is calculated from the same state discrimination measurement data, i.e. the detection matrix $\textbf{D}$. Although we study phase and amplitude tracking, a properly trained NN can in principle be used as an efficient high-dimensional estimator for tracking many sources of communication channel noise.

Figure \ref{nn_fig} shows a diagram of the NN architecture for the proposed noise tracking method, which has 10 layers (8 hidden), each with a Leaky ReLU activation function \cite{maas13}. To obtain the input for the NN, the detection matrix $\textbf{D}$ is first normalized across each row, and then arranged into a one-dimensional vector $(D_{l,k} \rightarrow D_{l(m+1) + k})$. This vector, along with the LO intensity for the previous $N$ measurements, are the inputs to the NN. For ease of notation, we denote the time-dependent input intensity of the QPSK coherent states as $\mathcal{A}(\tau) = |\alpha|^{2}(\tau)$ where $\tau$ represents time discretized into steps of $\Delta T$, where $1/\Delta T$ is the experimental repetition rate. For a single state discrimination measurement at time $\tau$, the intensity of the LO is denoted as $\mathcal{B}(\tau)=|\beta|^{2}(\tau)$. The NN output, denoted as $\hat{\mathcal{A}}_{NN}$ and $\hat{\phi}_{NN}$, are raw estimates of the input intensity $\mathcal{A}(\tau)$ and relative phase offset $\phi(\tau)$ during the previous $N$ state discrimination measurements. The NN is trained on $5\times10^{5}$ samples of the state discrimination measurement generated from Monte-Carlo simulations of the experiment in Python. For training the NN, we use the Tensorflow library \cite{abadi16} with a weighted mean squared error cost function (See Appendix A for details) \cite{sze17, jain96, goodfellow16, geron17}. The trained NN is then included in the Monte-Carlo simulations to perform parameter tracking on the state discrimination data such that the estimates from the NN are fed-forward to the LO to correct the measurement.
\section{Results}
We simulate the performance of the noise tracking method based on the NN estimator for a variety of scenarios with amplitude and phase noise for \textit{average} input intensities $\langle \hat{n} \rangle_{0} = \mathcal{A}(0) = \langle \mathcal{A}(\tau) \rangle$ equal to 2, 5, and 10. Here $\langle \cdot \rangle$ denotes the average across all noise realizations at the time step $\tau$. We benchmark the NN against a Bayesian estimator where the prior probability distribution for both parameters is uniform \cite{dimario20}. For all simulations, we use a single NN to perform multi-parameter estimation and noise tracking across a range of input powers and noise parameter regimes.

As a model for phase noise $\phi(\tau)$, we simulate a discrete Gaussian random walk in phase \cite{dimario20}. A single step of this walk has a variance of $\sigma_{1}^{2} = 2\pi \Delta\nu \Delta T$ where $\Delta\nu$ is the phase noise bandwidth due to finite laser linewidth \cite{lygagnon06,ip07,kikuchi08} or other phase noise sources \cite{kikuchi08,khanzadi15}. The experimental repetition rate is set to $1/\Delta T=100\mathrm{ MHz}$ such that $\Delta T = 10\mathrm{ ns}$ to represent a feasible, near-term communication bandwidth for non-Gaussian receivers \cite{holzman19}. To model amplitude noise of the input states, we simulate noise in the input intensity $\mathcal{A}(\tau)$. As a noise model, we use an Ornstein--Uhlenbeck (OU) process \cite{uhlenbeck30,gillespie96} whose stochastic differential equation is given by:
\begin{equation}
\Delta \mathcal{A}(\tau) = \gamma [\langle \hat{n} \rangle_{0} - \mathcal{A}(\tau)] \Delta T + \Sigma \sqrt{\Delta T} dW
\label{ou_eq}
\end{equation}
where $\gamma$ is the amplitude noise bandwidth, $\Sigma$ controls deviation of the walks, and $dW$ denotes a Wiener process. The long--time variance of $\mathcal{A}(\tau)$ is given by $\Sigma^{2}_\infty = \Sigma^{2}/2\gamma$ and the maximum long--time variance we implement is $\Sigma_{\infty}^{2}=\{0.25, 1.5, 6.0\}$ for $\langle \hat{n} \rangle_{0}=\{2, 5, 10\}$, respectively, corresponding to a relative noise level of $\langle \hat{n} \rangle_{0}/\Sigma_{\infty} \approx 0.25$.
\begin{figure}[!t]
	\includegraphics[width = 8.25cm]{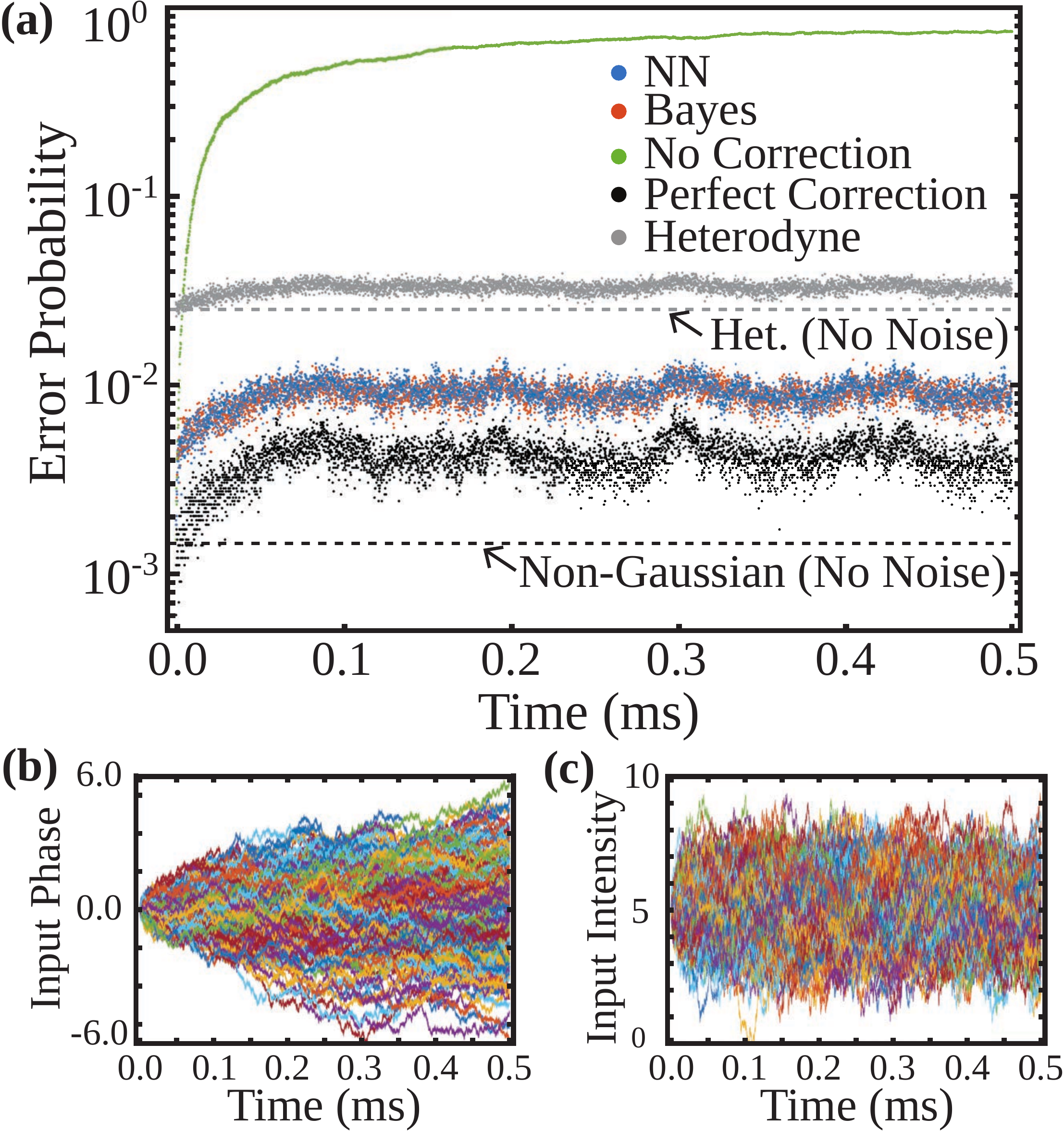}
	\caption{\textbf{Probability of error as a function of time}. (a) Error probability as a function of time for $\langle \hat{n} \rangle_{0}=5.0$ when both phase (b) and intensity (c) noise are applied and tracked. Here the noise parameters are: $\gamma=25$kHz, $\Sigma_{\infty}^{2}=1.5$, and $\Delta\nu=2$kHz. Blue (orange) points show the error for the NN (Bayes) based estimator. Green and black points show the error with no correction and perfect correction, respectively. Gray points show the effective QNL of a perfectly corrected heterodyne measurement. Black and gray dashed lines show the error for a non-Gaussian and heterodyne receiver, respectively, with no noise.}
	\label{mpn5_ex}
\end{figure}
\begin{figure*}[t]
	\includegraphics[width = 0.90\textwidth]{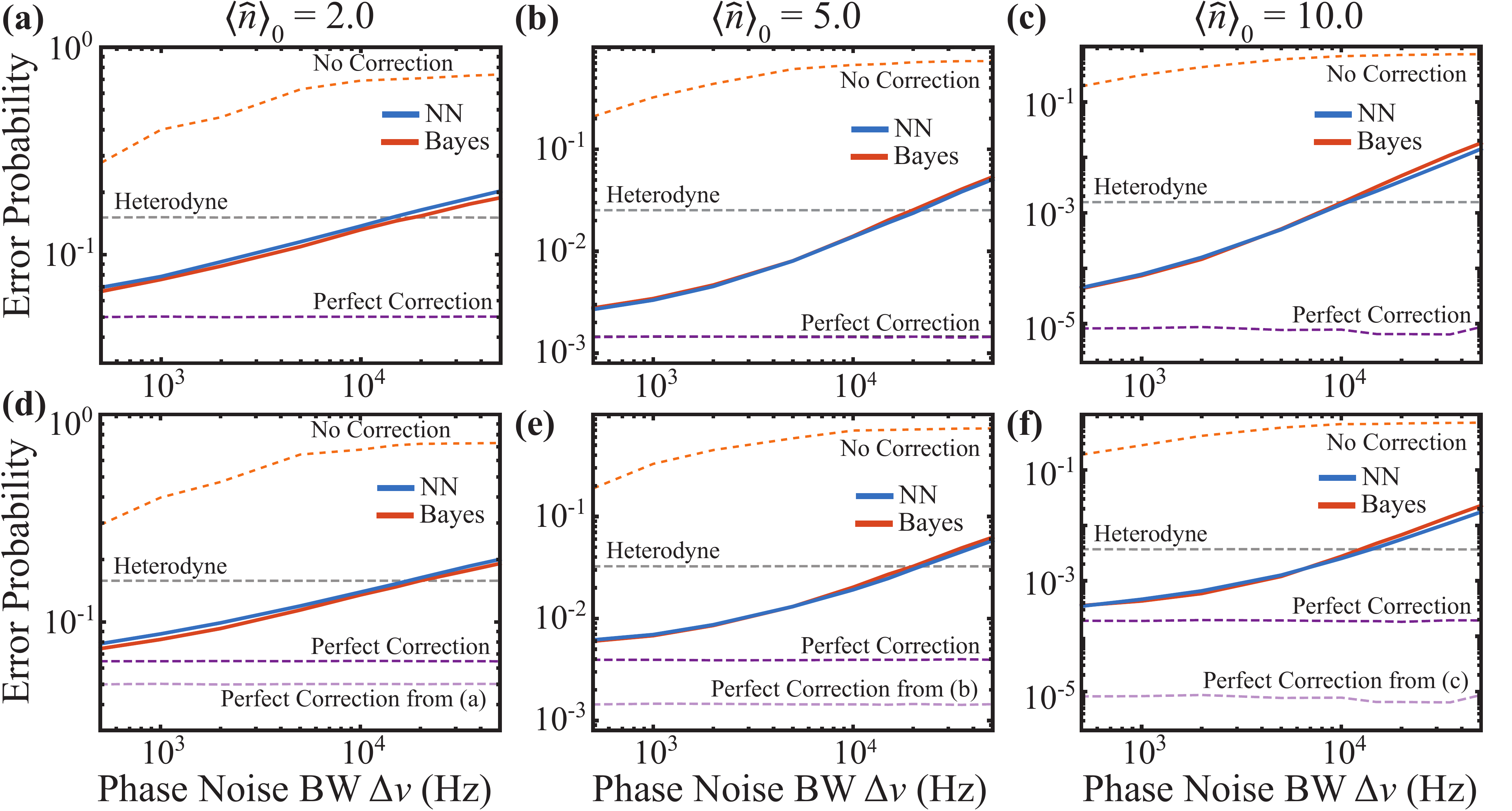}
	\caption{\textbf{Phase noise tracking.} Error probability as a function of phase noise bandwidth (BW) $\Delta\nu$ without (a)--(c) and with (d)--(f) amplitude noise of $\gamma=25$kHz and $\Sigma_{\infty}^{2}$ = 0.25, 1.5, and 6.0 for average intensities $\langle \hat{n} \rangle_{0}$ = 2, 5, and 10, respectively. Blue and orange lines show the performance of the noise tracking methods based on NN and Bayesian estimators, respectively. The orange dashed line shows the error probability for a non-Gaussian receiver with no correction. The purple and gray dashed lines show the error probability for a non-Gaussian and heterodyne measurement with perfect correction, respectively.}
	\label{lwscan}
\end{figure*}
After $N$ state discrimination measurements, estimates are calculated from the detection matrix \textbf{D}. To implement correction of the receiver, we set the LO intensity $\mathcal{B}(\tau)$ to the current estimated value $\hat{\mathcal{A}}(\tau)$ of the input intensity $\mathcal{A}(\tau)$. For phase tracking we add a correction $\delta(\tau)$ to the LO phase such that $\mathrm{arg}\{\beta\} = \theta_{j} + \delta(\tau)$. The correction $\delta(\tau)$ is equal to the cumulative sum of individual estimates $\hat{\phi}$ up to the current time step $\tau$. This is because the receiver is always estimating only the phase shift accumulated in the previous $N$ experiments. The phase and intensity corrections remain fixed at these values for $N$ experiments until new estimates are made and applied to the LO.

To reduce uncertainty in the phase and intensity estimates for noise tracking, we implement a Kalman filter \cite{kalman60} for both estimates (See Appendix B for details). The input for the filter are the current raw estimates for the input intensity and phase offset $(\hat{\mathcal{A}}_{NN}, \hat{\phi}_{NN})$, and the filter outputs are updated, filtered estimates for the intensity $\hat{\mathcal{A}}(\tau)$ and phase $\hat{\phi}$. The same procedure is done to obtain filtered Bayesian estimates from the raw estimates $(\hat{\mathcal{A}}_{B}, \hat{\phi}_{B})$. To implement the Kalman filter, we assume that the raw NN estimates are Gaussian distributed, and use Monte-Carlo simulations with fixed phase offset and input intensity to empirically obtain the variance of the NN estimator. We note that although we study two particular models for phase and amplitude noise, we believe this NN-based tracking method can be applied to a variety of noise forms such as power-law amplitude noise or damping noise. To study different noise models, the NN would need to be re-trained using data generated from the new model and the noise dynamics would need to be incorporated into the Kalman filter accordingly.
\begin{figure*}[t]
	\includegraphics[width = 0.90\textwidth]{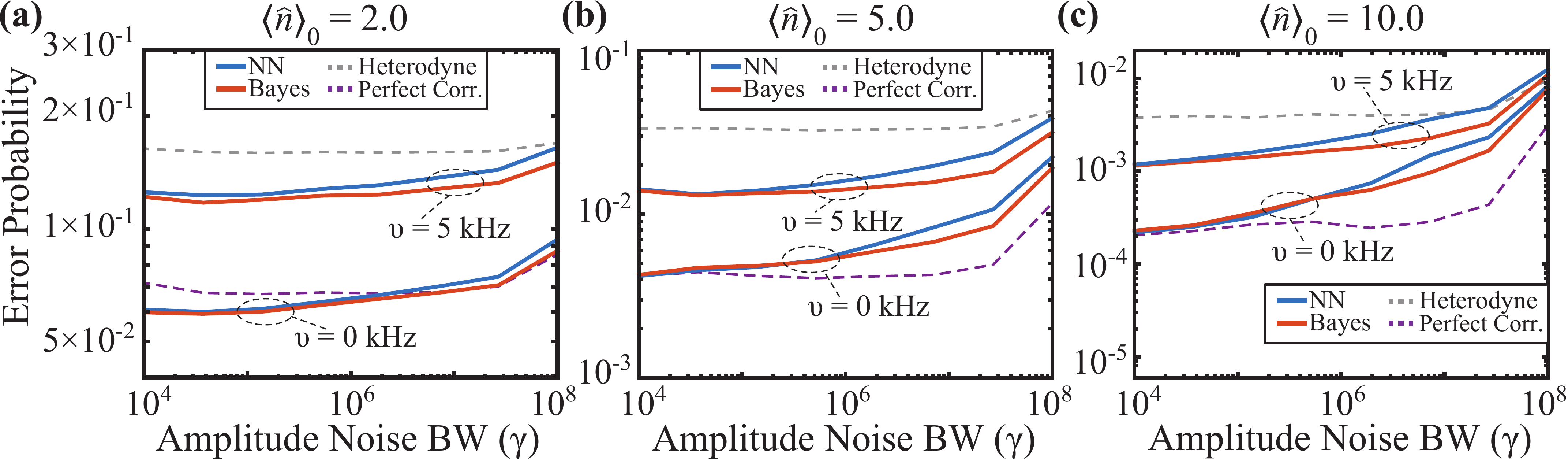}
	\caption{\textbf{Amplitude noise tracking.} Error probability as a function of amplitude noise bandwidth (BW) $\gamma$ for $\langle \hat{n} \rangle_{0}=\{2, 5, 10\}$ without and with phase noise with bandwidth $\Delta\nu=5\mathrm{kHz}$. Purple and gray dashed lines show the error for a non-Gaussian and heterodyne measurement with perfect correction, respectively. Beyond $\gamma\approx 10^{7}$ Hz, the amplitude noise is effectively random across the $N$ experiments.}
	\label{thscan}
\end{figure*}

Figure \ref{mpn5_ex} shows (a) the error probability of the non-Gaussian receiver with noise tracking for 1000 different realizations with both phase $(\Delta\nu=2 \mathrm{kHz})$, and intensity $(\gamma=25$kHz $, \Sigma_{\infty}^{2}=1.5)$ noise shown in (b) and (c), respectively, for an input intensity $\langle \hat{n} \rangle_{0}=5.0$ and $N=10$ experiments per estimation period. The blue (orange) points show the results the noise tracking method based on NN (Bayesian) estimators. The black points show the error probability with perfect correction, which corresponds to the case where the receiver has complete knowledge of the phase and intensity noise, so that $\mathcal{B}(\tau)=\mathcal{A}(\tau)$ and $\delta(\tau)=\phi(\tau)$. The green points show the error probability of an uncorrected non-Gaussian measurement, and the gray points show that of an ideal heteroydne measurement with perfect phase tracking (equivalent to $\phi(\tau)=0$). We note that even though the receiver may have perfect knowledge of the noise, the overall effect of the amplitude noise increases the error probability. This is because input powers smaller than the average power $(\mathcal{A}(\tau)<\langle \hat{n} \rangle_{0})$ increase the errors more than the reduction of error for larger powers $(\mathcal{A}(\tau)>\langle \hat{n} \rangle_{0})$. The dashed black and gray lines show the error for an adaptive non-Gaussian measurement, and an ideal heterodyne measurement with no noise, respectively. By comparing the error of the non-Gaussian measurement with perfect correction (black points) to the black dashed reference line, we observe that non-Gaussian measurements are more sensitive to amplitude noise than a heterodyne measurement (gray points vs gray dashed line), even when they are perfectly corrected. We observe that the NN-based tracking method performs equivalently to the Bayesian method, and both can allow the receiver to maintain an error probability significantly below the QNL. This result demonstrates the capabilities of a NN for efficient and reliable noise tracking for non-Gaussian receivers for state discrimination.

We study the robustness of the NN-based method in scenarios with different noise strengths and bandwidths in the phase and amplitude. For these studies, we use the heterodyne measurement with perfect phase tracking as the limit for conventional strategies, serving as the effective QNL when the same noise is applied to both receivers. In this section, we study (A) the error probability as a function of phase noise with fixed amplitude noise, and (B) the error probability when the amplitude noise levels are varied with phase noise with a fixed bandwidth.
\subsection{Phase noise with different bandwidths}

We study the performance of the noise tracking method based on the NN estimator as a function of the phase noise bandwidth $\Delta\nu$ for fixed values of amplitude noise $\Sigma_{\infty}^{2}$ and $\gamma$. We compare these results to the tracking method based on a Bayesian estimator, as well as a perfectly-corrected non-Gaussian measurement. We use different amplitude noise parameters for different values of $\langle \hat{n} \rangle_{0}$ such that the relative amplitude noise strength $\langle \hat{n} \rangle_{0}/\Sigma_{\infty}$ is constant. For an average intensity of $\langle \hat{n} \rangle_{0}=2, 5, $ and 10 we simulate 250, 250, and 500 different realizations of the noise, respectively. The simulations are run for $2\times10^{3}$ time bins of $N=10$ experiments each, giving a total of $2\times10^{4}$ individual experiments per noise realization. We calculate the average error across all realizations for all time bins.

Figure \ref{lwscan} shows the average error probability as a function of the bandwidth $\Delta\nu$ for intensities $\langle \hat{n} \rangle_{0}=$ 2, 5, and 10. Figures \ref{lwscan}(a)--\ref{lwscan}(c) have no amplitude noise ($\gamma=0$, $\Sigma^{2}_{\infty}=0$), and \ref{lwscan}(d)--\ref{lwscan}(f) have $\gamma=2$kHz, and $\Sigma_{\infty}^{2}=\{0.25, 1.5, 6.0\}$, respectively, corresponding to relative strength of $\langle \hat{n} \rangle_{0}/\Sigma_{\infty} \approx 0.25$. The performance of the noise tracking method based on the NN estimator (blue) is equivalent to the Bayesian-based method (orange) while being computationally efficient to implement. The purple and gray dashed lines show the average error of the non-Gaussian and ideal heterodyne receivers with perfect parameter tracking, respectively.

We observe that for all the investigated average input intensities $\langle \hat{n} \rangle_{0}$, the NN-based method performs as well as the Bayesian method both with and without amplitude noise. The NN-based method can enable the non-Gaussian receiver to surpass the QNL up to a phase noise bandwidth of $\Delta\nu\approx15$kHz, even in the presence of significant amplitude noise. We note that the situation in Figs. \ref{lwscan}(a)--\ref{lwscan}(c) with no amplitude noise is equivalent to the single parameter problem of phase tracking for non-Gaussian receivers as demonstrated in \cite{dimario20}. For large phase noise bandwidths for $\langle \hat{n} \rangle_{0}=5$, and 10, we observe that a NN estimator can perform slightly better than a Bayesian estimator. We believe this is due to the relatively small number of samples ($N=10$) from which estimates are made. In this regime with a few samples for estimation, there may be estimators which perform better than the Bayesian estimator, which is asymptotically optimal in the limit of many samples. Another potential cause of this effect is that in the training process of the NN, the relative weight between error in phase estimates and error in mean photon number estimates can be adjusted. This freedom may allow for fine tuning of the overall training error to allow for a slightly better overall performance, in terms of error probability, for specific channel models.
\subsection{Amplitude noise with different bandwidths}
To investigate the effect of the amplitude noise bandwidth $\gamma$, we fix the long-time variance $\Sigma_{\infty}^{2}$ and the phase noise bandwidth $\Delta\nu$. This allows for studying the performance of the NN-based method when the amplitude noise bandwidth $\gamma$ ranges from much smaller to much larger than the bandwidth for parameter estimation $1/N\Delta T$. Figure \ref{thscan} shows the average probability of error for different amplitude noise bandwidths $\gamma$ without and with phase noise of bandwidth $\Delta\nu=5$kHz, for $\langle \hat{n} \rangle_{0}=\{2, 5, 10\}$ with $\Sigma_{\infty}^{2}=\{0.25, 1.5, 6.0\}$, respectively. Blue (orange) lines show the error rates for the NN (Bayesian) based tracking method. Purple and gray dashed lines show the error probability for a non-Gaussian and heterodyne measurement with perfect correction, respectively. We find that the NN-based method performs closely to the Bayesian-based method, and enables the receiver to achieve sub-QNL error rates across a broad range of amplitude noise bandwidths even in the presence of phase noise.

We note that for intensity $\langle \hat{n} \rangle_{0}=2$ in Fig. \ref{thscan}(a), the error for both the noise tracking methods are below the perfectly corrected non-Gaussian measurement when $\Delta\nu=0$. At low input powers, strategies that optimize the LO intensity $(|\beta|^{2}>|\alpha|^{2})$ yield lower error probabilities than when $|\beta|^{2} = |\alpha|^{2}$ \cite{ferdinand17}. Due to the small number of samples $(N\times L)$ used for estimation, the NN and Bayesian estimators have a bias in $\hat{\mathcal{A}}_{B, NN}$, such that $\mathcal{B}(\tau)>\mathcal{A}(\tau)$. The effect of this bias in the intensity estimates $\hat{\mathcal{A}}_{B, NN}$ is that the corrected measurement unintentionally approximates an optimized strategy \cite{ferdinand17}. This effect results in error probabilities of the corrected receiver with both NN and Bayesian based methods that are below the error of a perfectly corrected nulling receiver where $\mathcal{B}(t)=\mathcal{A}(t)$, which is due to the bias of the estimators from finite sampling. Further investigation is needed to determine the capabilities of NN based noise tracking for optimized non-Gaussian receivers \cite{ferdinand17}.

The performance of the non-Gaussian receiver also depends on the long-time variance $\Sigma_{\infty}^{2}$ of the amplitude noise. In our main results, $\Sigma_{\infty}^{2}$ was set to represent a ``worst-case" scenario of $\approx 25\%$ relative amplitude noise (See Fig. \ref{mpn5_ex}). Appendix C describes our study of noise tracking of amplitude noise with different long-time variance $\Sigma_{\infty}^{2}$. In our findings, we observe that in the absence of phase noise, both the NN and Bayesian-based tracking methods enable the receiver to perform below the QNL, and close to the performance of perfect noise correction. In the presence of phase noise with bandwidth $\Delta\nu=5$kHz, the sub-QNL performance of the receiver is maintained, and the effect of increasing $\Sigma_{\infty}^{2}$ is small compared to the effects of increasing phase or amplitude bandwidths.
\section{Discussion}
The numerical studies in this work show that methods for channel noise tracking based on NN estimators are able to accurately track dynamic phase and amplitude noise to allow an adaptive non-Gaussian measurement to maintain performance below the QNL. We note that in the asymptotic limit of many samples available for parameter estimation for noise tracking, a Bayesian estimator can achieve minimal mean-square error \cite{lehmann98}. However, when noise tracking and correction need to be realized in real time to reduce errors in the state discrimination measurement and generate reliable data for parameter estimation, there is always a limited number of samples from which estimates are made. In these situations, there is a trade-off between estimation precision and noise tracking bandwidth. Other estimators, such as a NN, may balance these two parameters better than a Bayesian estimator for increased precision with finite samples. This property can enable efficient methods for high dimensional parameter tracking of complex dynamic channel noise.

The computational efficiency of the NN estimator is rooted in the small number of multiplications required to calculate a single estimate, the limited memory requirements, and in the fact that the NN method does not explicitly depend on the value of $N$ or the number of adaptive steps $L$, as opposed to the Bayesian approach. For example, the NN-based estimator in this work requires $\approx5500$ multiplications. On the other hand, a Bayesian estimator using likelihood functions which are discretized into a $100 \times 100$ grid would require $100^{2}\times N \times L = 10^{6}$ multiplications, which may not be compatible with devices such as FPGAs \cite{becerra15,dimario20}. While there may be methods to reduce this computational cost, the Bayesian estimator also would require storage in memory of the full photon counting likelihood functions, putting stringent requirements on the device memory. For example, the $100\times100$ grid for the Bayesian estimator with 16 bit precision would require 800 kB of memory simply to store the likelihood functions. Moreover, to extend the noise tracking method for estimation of three noise parameters, a NN would simply require a single added output and proper retraining, while a Bayesian estimator would require possibly $10^{8}$ multiplications and 80 MB of memory.

The robustness and versatility of the NN-based noise tracking method described here, shows that NN-based methods can be practical and very useful tools for non-Gaussian receivers. In addition, other machine learning techniques, such a reinforcement learning \cite{bilkis20, foesel18}, could provide a further benefits to these non-conventional measurements when the best detection strategy may be unknown or infeasible to calculate. We anticipate that neural networks and machine learning will have a great benefit for non-Gaussian measurements, just as these techniques have proven worthwhile for conventional measurement strategies \cite{thrane17, wu09, zibar15, zibar16, khan17, wang17, lohani19, karanov18}.
\section{Conclusion}
We investigate the use of a neural network (NN) as a computationally efficient multi-parameter estimator of dynamic channel noise, enabling robust noise tracking for adaptive non-Gaussian measurements for coherent state discrimination. We study the NN-based tracking method for simultaneous amplitude and phase noise and find that the NN estimator can perform as well as a more complex Bayesian estimator. This performance is observed across a broad range of noise strengths and bandwidths for different average powers of the input coherent states. The non-Gaussian receiver used in this study can have broad applications in classical \cite{lee16,becerra15} and quantum communication \cite{liao18,guan16} due to its ability to attain sensitivities beyond the quantum noise limit (QNL). Moreover, the proposed method for noise tracking uses only the data collected during the state discrimination measurement without requiring extra resources such as strong reference pulses. This makes the receiver and the proposed method for noise tracking well suited for energy efficient low power communications. Thus, NN based methods are ideal candidates for real-time tracking of multiple sources of channel noise for non-Gaussian receivers, allowing them to maintain their sub-QNL sensitivity in the presence of complex dynamic channel noise.

\begin{acknowledgments}
This work was supported by the National Science Foundation (NSF) (PHY-1653670, PHY-1521016, PHY-1630114). The source code is available at: github.com/UNM-QOlab/phase\_amp\_tracking\_nn
\end{acknowledgments}

\appendix

\section{Neural network estimator training}
\begin {table}[!b]
\begin{center}
	\caption{Neural Network and training parameters.}
	\begin{tabular}{ |c|c| }
		\hline
		\multicolumn{2}{|c|}{Network parameters}\\
		\hline
		Number of Layers & 10 (8 Hidden)\\
		Hidden layer size $(l_{i})$ & $\{32, 32, 32, 32, 16, 16, 8, 8\}$\\
		Activation function & Leaky ReLU ($a = 0.1$)\\
		Initialization & Norm(0, $\sigma^{2}=1/l_{i}$) (Xavier)\\
		\hline
		\multicolumn{2}{|c|}{Training parameters}\\
		\hline
		Cost function & Weighted mean squared error\\
		Optimizer & RMSprop (momentum=0.8)\\
		Learning rate & $50\times 10^{-6}$ \\
		Epochs & 2000\\
		\hline
	\end{tabular}
	\label{nn_param}
\end{center}
\end {table}
To generate training data, we use Monte-Carlo simulations of the adaptive non-Gaussian measurement described in Section II(A) \cite{becerra15}. For a single training data element, the strategy is simulated using a randomly chosen intensity for the input state $\mathcal{A}$ and LO intensity $\mathcal{B}$, both sampled from a uniform distribution $\mathcal{U}(0.05, 25.0)$. A constant value for the phase noise $\phi$ is then sampled from a Gaussian distribution $\mathcal{N}(0, \sigma=0.25)$. In addition, we randomly sample the number of experiments $N$ that comprise each sample from a uniform distribution $\mathcal{U}(2, 200)$. This random sampling of the three parameters enables the NN to be used for different values of $N$ depending on the noise characteristics. During the training of the NN, we use true values of the input parameters $(\phi, \mathcal{A})$ as the target values. This procedure using random sampling of multiple input noise parameters ensures sufficient sampling of the input parameter space, enabling training of a robust NN estimator

To train the NN we use the Tensorflow framework \cite{abadi16} in Python. Table \ref{nn_param} summarizes the relevant NN and training parameters. We use the RMSprop optimizer \cite{tieleman12} with a weighted mean-squared-error cost function. The weight $w_{i}$ of each training sample is given by:
\begin{equation}
w_{i} = e^{-(\mathcal{A}_{i} - \mathcal{B}_{i})^{2} / 2} + 0.1
\end{equation}
where $\mathcal{A}_{i}$, and $\mathcal{B}_{i}$ are the intensity of the input states and LO of the $i^{th}$ sample, respectively. This allows the NN to accurately estimate the parameters when the LO and input intensities are close to each other, as one would expect in practice, while also being somewhat robust to large amplitude fluctuations.

\section{Kalman filter}
We implement Kalman filtering \cite{kalman60, welch00} of both the phase estimate and intensity estimate in order to reduce the uncertainty in the applied corrections to the adaptive non-Gaussian measurement. For the phase estimates, the predicted mean value $\hat{y}_{\phi}$ and variance $\hat{\sigma}_{\phi}^{2}$ are:
\begin{align}
\hat{y}_{\phi} &= 0
\\
\nonumber
\hat{\sigma}_{\phi}^{2} &= \sigma_{\phi}^{2} + N\sigma_{1}^{2}
\end{align}
where $\hat{y_{\phi}}$ represents the predicted average value for the phase, $\sigma_{\phi}^{2}$ the variance of the current prior probability distribution for the phase, $\sigma^{2}_{1}=2\pi\Delta\nu\Delta T$,  and $\hat{\sigma}_{\phi}^{2}$ the predicted phase variance. The filtered estimate $\hat{\phi}$ is then obtained from the raw estimate $\hat{\phi}_{NN}$ and (B1) by:
\begin{align}
\hat{\phi} &= K_{\phi}\hat{\phi}_{NN} + (1-K_{\phi})\hat{y}_{\phi}
\\
\nonumber
\sigma_{\phi}^{2} &= (1-K_{\phi}) \hat{\sigma}_{\phi}^{2}
\\
\nonumber
K_{\phi} &=\frac{\hat{\sigma}_{\phi}^{2}}{\hat{\sigma}_{\phi}^{2} + \sigma_{\phi, NN}^{2}}
\end{align}
where $K_{\phi}$ is the Kalman gain for the phase estimate, $\sigma^{2}_{\phi, NN}$ is the variance of the NN phase estimate, and $\sigma^{2}_{\phi}$ is the updated variance of the filtered phase estimate.

Similarly, the equations for the predicted input intensity mean value $\hat{y}_{\mathcal{A}}$ and variance $\hat{\sigma}_{\mathcal{A}}^{2}$ are given by:
\begin{align}
\hat{y}_{\mathcal{A}} &= (1-\gamma \Delta T)^{N}\mathcal{B}_{N} + \gamma \langle \hat{n} \rangle_{0} \Delta T\sum\limits_{k=0}^{N-1}(1-\gamma \Delta T)^{k}
\\
\nonumber
\hat{\sigma}_{\mathcal{A}}^{2} &= (1-\gamma \Delta T)^{2N}\sigma_{\mathcal{A}}^{2} + \Sigma^{2} \Delta T\sum\limits_{k=0}^{N-1}(1-\gamma \Delta T)^{2k}
\end{align}
where $\gamma$ is the amplitude noise bandwidth, $\langle \hat{n} \rangle_{0}$ is the average input intensity equal to $\mathcal{A}(0)$, and $\mathcal{B}_{N}$ is the LO intensity for the previous $N$ experiments. These equations result from propagating the mean and variance of Eq. (\ref{ou_eq}) for $N$ time steps of duration $\Delta T$.

The filtered intensity estimate $\hat{\mathcal{A}}$ and variance $\sigma_{\mathcal{A}}^{2}$ are then obtained from the raw estimate $\hat{\mathcal{A}}_{NN}$ and (B3) by:

\begin{align}
\hat{\mathcal{A}} &= K_{\mathcal{A}}\hat{\mathcal{A}}_{NN} + (1-K_{\mathcal{A}})\hat{y}_{\mathcal{A}}
\\
\nonumber
\sigma_{\mathcal{A}}^{2} &= (1-K_{\mathcal{A}}) \hat{\sigma}_{\mathcal{A}}^{2}
\\
\nonumber
K_{\mathcal{A}} &=\frac{\hat{\sigma}_{\mathcal{A}}^{2}}{\hat{\sigma}_{\mathcal{A}}^{2} + \sigma_{\mathcal{A}, NN}^{2}}
\end{align}
where $K_{\mathcal{A}}$ is the Kalman gain for the intensity estimate and $\sigma_{\mathcal{A}, NN}^{2}$ is the variance of the NN phase estimate. The initial $(\tau=0)$ variances for both phase and intensity are set to zero.

\begin{algorithm}[t]
	\caption{NN Parameter Tracking Algorithm}
	\begin{algorithmic}
		\Function {KalmanFilter}{$\hat{A}_{NN}, \hat{\phi}_{NN}$}
		
			\State Predict: $\hat{y}_{\phi}, \hat{\sigma}^{2}_{\phi}, \hat{y}_{\mathcal{A}}, \hat{\sigma}^{2}_{\mathcal{A}}$ \Comment{Eqs. (B1), (B3)}
			
			\State $\hat{\phi} \leftarrow K_{\phi}\hat{\phi}_{NN} + (1-K_{\phi})\hat{y}_{\phi}$ \Comment{Eq. (B2)}
			
			\State $\hat{\mathcal{A}}(\tau) \leftarrow K_{\mathcal{A}}\hat{\mathcal{A}}_{NN} + (1-K_{\mathcal{A}})\hat{y}_{\mathcal{A}}$ \Comment{Eq. (B4)}
			
			\State Update: $K_{\phi}, \sigma_{\phi}^{2}, K_{\mathcal{A}}, \sigma_{\mathcal{A}}^{2} $ \Comment{Eqs. (B2), (B4)}
			
			\State \textbf{return} $\hat{\phi}, \hat{\mathcal{A}}(\tau)$
			
		\EndFunction
		\\
		\State \textbf{Initial}
		\State $\mathcal{B}(0) \leftarrow \hat{\mathcal{A}}(0)$
		\State $\delta(0) \leftarrow 0$
		\State $\tau \leftarrow 0$ \Comment Time in increments of symbol time
		\State $n \leftarrow 0$ \Comment Number of measurements performed
		\Loop
			\State $\{d_{j}\}_{L}, \{\Delta_{j}\}_{L}$ $\leftarrow$ Single discrimination measurement
			\State Add $\{d_{j}\}_{L}, \{\Delta_{j}\}_{L}$ to detection matrix \textbf{D}
			\State $\tau \leftarrow \tau + \Delta T$
			\State $n \leftarrow n + 1$
			\If{$n=N$} \Comment{Update LO every $N$ measurements}
				
				\State $\hat{A}_{NN}, \hat{\phi}_{NN}$ $\leftarrow$ Evaluate NN with inputs \textbf{D}, $\mathcal{B}(\tau)$
				\State Reset \textbf{D} to zeros
				\State $\hat{\phi}, \hat{\mathcal{A}}(\tau) \leftarrow$ \textsc{KalmanFilter}($\hat{A}_{NN}, \hat{\phi}_{NN}$)
				\State $\mathcal{B}(\tau) \leftarrow \hat{\mathcal{A}}(\tau)$ \Comment{Correct LO intensity}
				\State $\delta(\tau) \leftarrow \delta(\tau) + \hat{\phi}$ \Comment{Add estimate to phase correction}
				\State $n \leftarrow 0$ \Comment{Reset measurement counter}
			\Else \Comment{Don't update LO}
				\State $\mathcal{B}(\tau) \leftarrow \mathcal{B}(\tau-\Delta T)$ 
				\State $\delta(\tau) \leftarrow \delta(\tau-\Delta T)$
			\EndIf
		\EndLoop
		
	\end{algorithmic}
	\label{algo}
	
\end{algorithm}
In order to empirically obtain the variances $\sigma_{\phi, NN}^{2}$, and $\sigma_{\mathcal{A}, NN}^{2}$ we use Monte Carlo simulations of the experiment and the NN estimator without the Kalman filter. For all average intensities, we fix the input intensity to $\mathcal{A}(t)=\{2, 5, 10\}$ and phase noise to zero, and calculate the variance of $10^6$ estimates. The variance is calculated for a range of the number of experiments per estimation $N$. We fit the variance as a function of $N$ to a power law function so that the filter may be used for different values of $N$. For $N=10$, as used in the simulations, $\sigma_{\mathcal{A}, NN}^{2}=\{3.37, 5.63, 5.35\}\times10^{-1}$ and $\sigma_{\phi, NN}^{2}=\{4.77, 3.61, 2.66\}\times10^{-3}$ for $\langle \hat{n} \rangle_{0}=\{2, 5, 10\}$, respectively. As discussed in Section \textbf{IV B}, the NN estimator is not necessarily unbiased across the range of parameters it is trained on due to the small sample size of only $N\times L$ as well as imperfections in the NN training process.

Employing a Kalman filter allows for construction of the full NN-based parameter tracking method. Algorithm \ref{algo} shows the pseudo-code for running the NN-based method including the filtering steps. For every time step $\tau$, a single state discrimination measurement is completed, which yields detections $\{d_{j}\}_{L}$ and relative phases $\{\Delta_{j}\}_{L}$ which populate the detection matrix $D$, as described in Section IIB. Every $N$ time steps, i.e. every $N$ measurements, the NN is evaluated to provide raw estimates $\hat{\mathcal{A}}_{NN}, \hat{\phi}_{NN}$ of the intensity and phase offset within the previous $N$ measurements. These raw estimates are then passed through the Kalman filter, which returns the current, filtered estimates for the intensity $\hat{\mathcal{A}}(\tau)$ and phase offset $\hat{\phi}$. These current estimates are then used to update the LO parameters $\mathcal{B}(\tau),\delta(\tau)$ for the next $N$ state discrimination measurements.

\begin{figure}[!t]
	\includegraphics[width = 8.25cm]{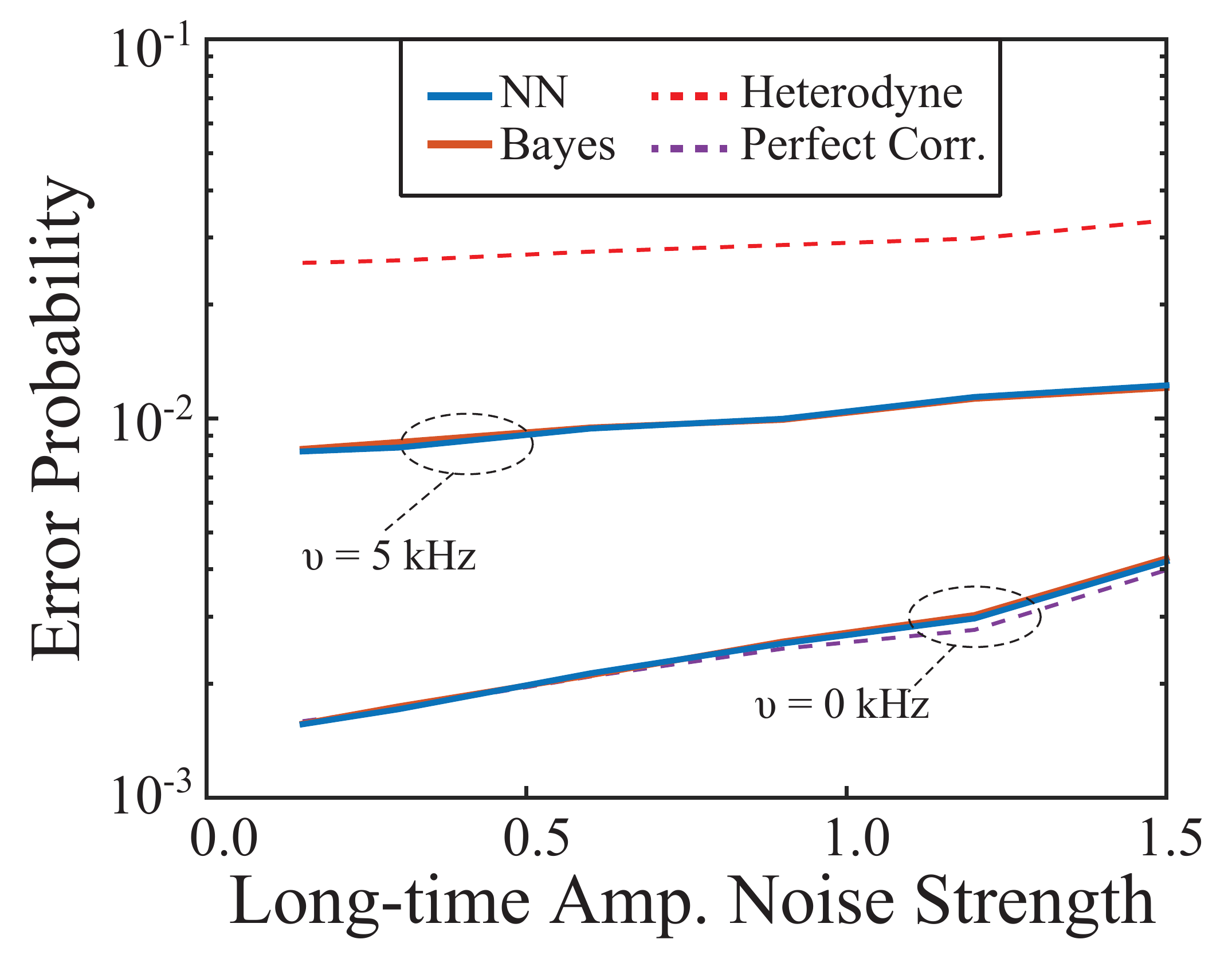}
	\caption{Error probability as a function of the long time variance $\Sigma_{\infty}^{2}$ of the amplitude noise for $\langle \hat{n} \rangle_{0}=5$, $\gamma=25$kHz without and with phase noise of bandwidth $\Delta\nu=5$kHz.}
	\label{mpnstrscan}
\end{figure}
\section{Different magnitudes of amplitude noise}	
We further investigate the performance of the NN estimator when varying the long-time strength of the amplitude noise $\Sigma_{\infty}^{2}$. Figure \ref{mpnstrscan} shows the error probability of the receiver for $\langle \hat{n} \rangle_{0}=5$ across a range of $\Sigma_{\infty}^{2}$ for fixed $\gamma=25$kHz without and with phase noise of bandwidth $\Delta\nu=5$kHz. We find that the NN based tracking method performs similar to the one based on the Bayesian estimator, and enables an error probability below that of the ideal heterodyne measurement. We have observed in other studies that the behavior for average input intensities $\langle \hat{n} \rangle_{0}=$2 and 10 is similar to the one for $\langle \hat{n} \rangle_{0}=$5 in Fig. \ref{mpnstrscan}.
\begin{figure}[!t]
	\includegraphics[width = 8.25cm]{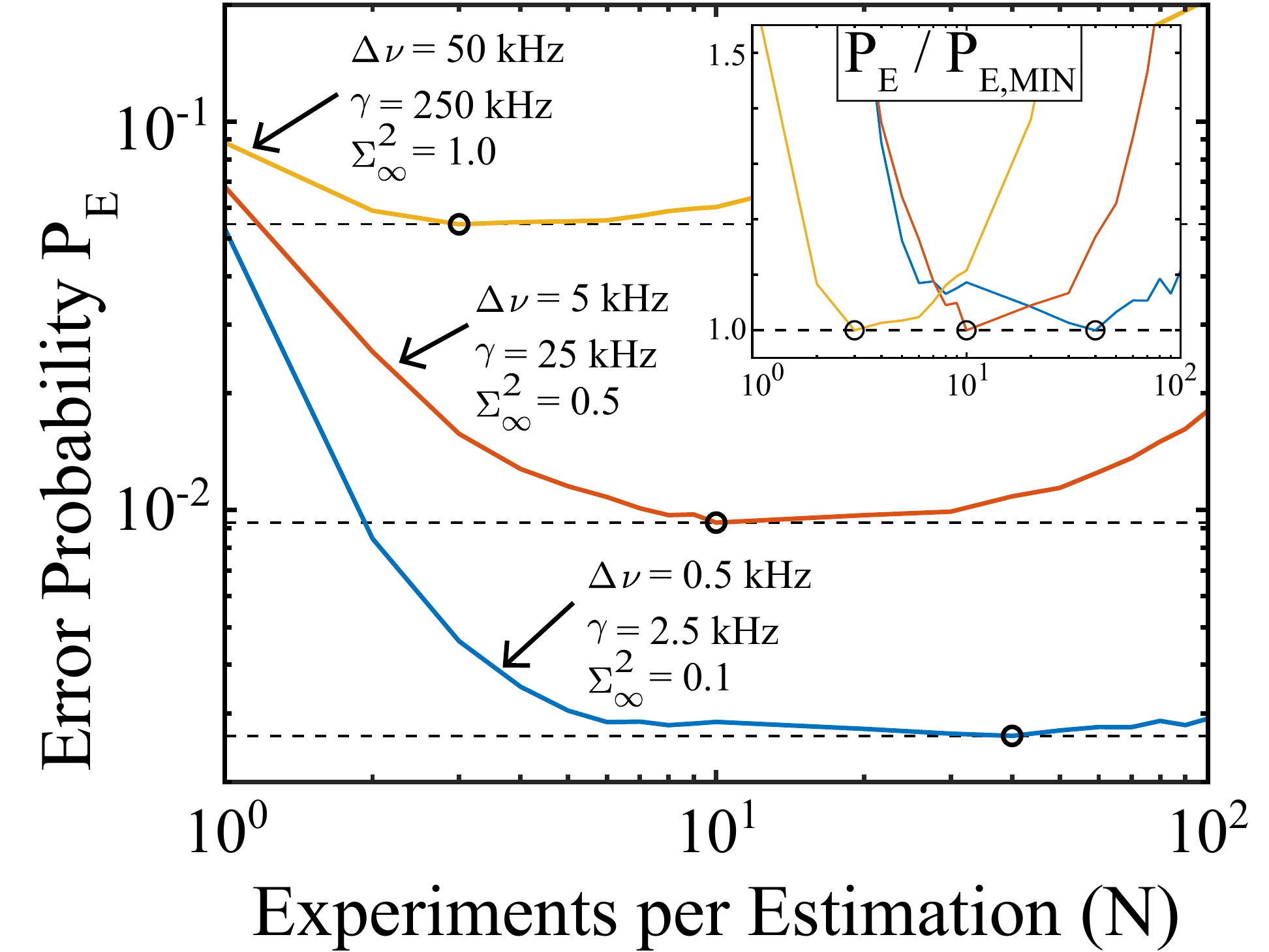}
	\caption{Error probability as a function of $N$ for three different noise strengths: low noise (blue), moderate noise (orange), and high noise (yellow). The black circles show the minimum error probability $\mathrm{P}_{E, MIN}$ and corresponding optimal value for N for each noise regime. The inset shows the error probability normalized by the minimum error given by the black circles in the main figure.}
	\label{estvar}
\end{figure}
\section{Estimation Time-Bandwidth Trade-off}
The overall performance of the phase tracking also depends on the number of experiments $N$ used to obtain a single estimate. For this study, we fixed $N$=10 for all simulations to demonstrate the versatility of the NN estimator. Although, realistic implementations may use a particular channel with specific noise characteristics. In this scenario, the value of $N$ can be fine tuned to optimize the performance of the phase tracking method to balance the estimation bandwidth (smaller $N$) and estimation accuracy (larger $N$). The optimization aims to find a value of $N$ such that the overall error probability is minimized when implementing the noise tracking method. The Kalman filtering attempts to balance the effects of the estimation variance and the noise variance over $N$ measurements in an optimal way through the Kalman gain $K$. However, there is still a trade-off between these two variances and a value of $N$ which achieves minimal $\mathrm{P}_{E}$ for specific channel noise conditions.

Figure \ref{estvar} shows the overall error probability as a function of $N$ when implementing the NN estimator with filtering for three different sets of channel noise parameters for $\langle \hat{n} \rangle_{0}=5.0$. The blue line corresponds to $\Delta\nu=0.5$kHz, $\gamma=2.5$kHz, and $\Sigma_{\infty}^{2}=0.1$. The orange line corresponds to $\Delta\nu=5$kHz, $\gamma=25$kHz, and $\Sigma_{\infty}^{2}=0.5$. The yellow line corresponds to $\Delta\nu=50$kHz, $\gamma=250$kHz, and $\Sigma_{\infty}^{2}=1.0$. For small noise bandwidth and strength (blue), the optimal value of $N$ (black circles) is approximately $N=40$, but decreases to $N=10$ and $N=3$ as the noise bandwidth and strength increase (orange and yellow). The inset shows the error probability normalized by the minimum for each noise strength for clarity. These optimal values of $N$ for specific channel noise parameters represent the optimal balance between estimation uncertainty and accumulated uncertainty from the channel noise. Thus, for a channel with known noise characteristics, an optimal value of $N$ can be found. In the studies presented in the main manuscript, we fixed $N$=10, since we found that this value allows the receiver to be versatile and operate well across a wide range of noise bandwidths. In the inset, $N=10$ is optimal for moderate noise levels. For small and large noise levels, the error at $N=10$ is only slightly higher compared to their respective minimums. Thus, for a specific well known channel an optimal $N$ can be implemented, but for a robust and versatile implementation a different value of $N$ may be beneficial.

\end{document}